\documentstyle[twocolumn,aps,prl,epsf]{revtex}
\tighten
\draft

\def\ha{Hamiltonian }
\def\<{\langle}
\def\>{\rangle}
\def\sus{susceptibility }

\def\half{{\frac{1}{2}}}
\def\be{\begin{equation}}
\def\ee{\end{equation}}
\def\imchi{\chi^{\prime \prime}(\nu)}

\begin{document}
\preprint{cond-mat}
\title{Localization transition in the Mermin model}

\author{Gregory Levine}

\address{Department of Physics, Hofstra University, Hempstead, NY 11550\\and}
\address{Department of Physics, Brookhaven National Laboratory, 
Upton, NY 11973-5000}

\author{V. N. Muthukumar}
\address{Joseph Henry Laboratories of Physics, Princeton University, 
Princeton, NJ 08544}
\date{\today}
\maketitle
\begin{abstract}
We study the dynamical properties of the Mermin model, a simple
quantum dissipative model with a monochromatic environment, using
analytical and numerical methods. Our numerical results show that the
model exhibits a second order phase transition to a localized state
before which the system is effectively decoupled from the environment.
In contrast to the spin-boson model, the Mermin model exhibits an
``orthogonality catastrophe,'' defining the critical point, before
dissipation has destroyed all coherent behavior. An analytic approach
based on the Liouvillian technique, though successful in describing
the phase diagram of spin-boson and related models, fails to capture
this essential feature of the Mermin model.

\end{abstract}


\section{Introduction}
\label{intro}

The formal analogies between phase transitions in one-dimensional
systems, decoherence in the spin-boson Hamiltonian (SBH),
and screening in
Kondo systems have been important in sorting out the subtle behavior
of dissipative quantum systems.  In this paper we undertake an
analytic and numerical study of the Mermin model, a model closely
related to the spin-boson model, but possessing a monochromatic
environmental spin bath.  

The spin-boson model is a universal model for a
two level system interacting weakly with an environmental bath, 
which has a continuous spectrum \cite{leggett_rmp}.  
Although both monochromatic and
polychromatic baths bring about a localization transition in the two
level system, we point out, in this
paper, an important difference between the dynamics of the
two level system interacting weakly with these two kinds of baths.  If
the dynamics of the two level system is characterized by a (pseudo) spin 
susceptibility, increasing the coupling to
the environment broadens the resonance and shifts the
resonant frequency somewhat. When the environment has a continuous and
ohmic spectrum (the SBH), the system spin first
loses all coherent behavior at some intermediate coupling strength.
The dynamics of the two level system becomes overdamped at this point,
but the localization transition does {\em not} occur
until a stronger coupling is reached.
The critical coupling at which the two level system is 
ultimately localized
may be identified with an ``orthogonality
catastrophe.''  Thus, before the environmental overlap effects become
catastrophic, they are {\sl finite} and lead to a continuous
softening of the system energy scale, as well as a complete inelastic
broadening to an incoherent state (the so called ``Toulouse point''). 

On the other hand,
when the spectrum of the environment is monochromatic (the Mermin
model), the resulting dynamics are different. As the coupling to the
environment increases, the only effect is to broaden
the coherent resonance.
At a critical coupling, a redistribution of spectral weight 
is initiated and
spectral weight flows from the coherent resonance to a localized feature
at zero frequency. For coupling strengths beyond this value, spectral
weight is increasingly transferred from the resonant peak 
to the localized feature. Thus, unlike the SBH, the Mermin model does
not show any overdamped feature before the localization transition sets
in. When localization does set in (at the critical point), 
{\em the coherent peak continues to exist, albeit with reduced 
spectral weight}. This is our main result.

The paper is organized as follows. In section \ref{an_res}, we 
introduce the Mermin model and study the dynamics of the model using
an analytical technique. This technique (based on Liouvillian operator
methods) has been applied successfully in the past, to a variety of
dissipative models, including the SBH. In section \ref{num_res}, we
study the model numerically, and show that the model exhibits a
localization transition at a critical 
value of the coupling strength. The final section
contains some concluding remarks, summarizing our results and
suggesting future directions of study.


\section{Liouvillian dynamics of the Mermin model}
\label{an_res}

In this section, we introduce the Mermin model and study the dynamics of
the model analytically. The model introduced by Mermin is
obtained from the well known spin-boson
Hamiltonian, by retaining only the lowest two levels of each
harmonic oscillator comprising the environmental bath
\cite{mermin91}. The \ha therefore describes a ``system'' spin-1/2,
represented by the Pauli matrices $\bf \sigma$, coupled to a set of
$N$ ``environment'' spin-1/2 degrees of freedom, $\{{\bf s}_j\}$. The
resulting Hamiltonian, called the Mermin model is given by 
\be
\label{ham}
H=-\frac{\Delta}{2} \sigma_x + \frac{\lambda}{4N} \sigma_z
\sum_{j=1}^{N}{(s_j^+ + s_j^-)} + \frac{\omega}{2N}
\sum_{j=1}^{N}{s_j^z}~~. 
\ee 

In this paper, we investigate a large $N$ variant of the model
(\ref{ham}). As first noted by Mermin, the model can be solved
approximately in the $N \rightarrow \infty$ limit, to demonstrate the
correspondence between the onset of localization of the system spin
and a second order phase transition \cite{mermin91}.  In this large
$N$ model, the environment spins are summed to one big $O(N)$ spin,
$\bf S$, and contributions from the sectors of Hilbert space with
total spin $S < N/2$ are ignored.  The \ha then becomes \be
\label{ham2}
H=-\frac{\Delta}{2} \sigma_x + \frac{\lambda}{2N} \sigma_z S_x +
\frac{\omega}{2N} S_z~~. 
\ee 

As Mermin points out, the advantage of the \ha (\ref{ham2}) is that in
the limit $N \rightarrow \infty$, the environment spins may be
replaced in the \ha by the $x$ and $z$ components of a classical spin
angular momentum: $\frac{\lambda}{N} \sigma_z S_x \rightarrow \half
\lambda \sigma_z \sin{\theta}$ and $\frac{\omega}{N} S_z \rightarrow
-\half \omega \cos{\theta}$.  The resulting Hamiltonian may be
diagonalized and its ground state eigenvalue, $E_0(\theta)$, given by
\be 
E_0(\theta) = -\half \sqrt{\Delta^2 + \frac{\lambda^2}{4}
\sin^2{\theta}} - \frac{\omega}{4} \cos{\theta}~~,
\ee 
minimized with respect to $\theta$.

The critical behavior of the model may now be seen by examining the
ground state energy $E_0$, as a function of $\Delta$. 
The ground state energy, $E_0$, bifurcates at a
finite value of $\Delta$ going from a singlet, non-degenerate root for
$\Delta >\lambda^2/2\omega$ to a doubly degenerate set for
$\Delta <\lambda^2/2\omega$:
\begin{eqnarray}
\theta_0 = 0~~~~~~~~~~~~~~~~~~~~~~~~~~~ &  \Delta>\lambda^2/2\omega 
\nonumber \\
\sin{\theta_0} = \pm \sqrt{\frac{1-4\Delta^2\omega^2/\lambda^4}
{1+\omega^2/\lambda^2}}~~~~~~ &  \Delta<\lambda^2/2\omega \nonumber
\end{eqnarray}
In the former case, the environment is decoupled from the system and
always points along the $z$-axis to minimize its ``Zeeman'' energy
corresponding to the last term of (\ref{ham2}).  In the latter case, the
environment and the system spins are frozen in two distinct
orientations with degenerate energies.  The ground state wave function
is still in a superposition $\alpha|+\>+\beta|-\>$, but is no longer an
evenly weighted one ($\alpha=\beta$); rather, the two roots correspond
to the system predominantly in  $|+\>$ or $|-\>$.

Despite producing a localization transition, this mean field solution
is necessarily incomplete, for, it ignores quantum fluctuations of the
bath; {\em i.e.}, it provides no mechanism for dissipation. As a first
step in this direction, we study the dynamics of the bath using the
Liouvillian operator method.  This method has been applied
successfully to study a wide variety of dissipative systems.  Using
the Liouvillian approach, Shao and H\"anggi \cite{hanggi98} studied a
spin-spin bath system similar to the Hamiltonian (\ref{ham}) above,
but with an ohmic spectrum and found a phase diagram virtually
identical to that of the SBH.  Dattagupta et al \cite{dattagupta89},
studying the SBH, also employed a Liouvillian approach and found close
agreement with other analytic approaches to the SBH
\cite{leggett_rmp,weiss_book}.

To try to gain some insight into the Mermin model beyond the mean
field level, we have adapted the resolvent operator calculation of
Shao and H\"anggi \cite{hanggi98} on the spin-spin bath model. 
In what follows, we just outline the basic steps in the
calculation and describe our results. A detailed review of the
resolvent operator formalism can be found elsewhere 
\cite{dattagupta_book}.

The Hamiltonian (\ref{ham2}) is rotated by $R
\equiv \exp{-i \sigma_z \theta_0 S_y}$ to diagonalize the environment
spin for the two orientations of the system spin. We then get
\be
\label{ham3}
H^{\prime} = -\frac{\Delta}{2} (s^+ e^{-2i \theta_0 S_y} + 
s^- e^{+2i \theta_0 S_y}) - \frac{(\omega^2 + \lambda^2)^{1/2}}{2N}
S_z~~.
\ee

To study the dynamics of the system spin, a suitable quantity to
calculate is $\langle \sigma_z(t) \rangle$ where $\langle \ldots
\rangle$ denotes quantum average at zero temperature.  For a state
prepared initially at $t=0$, with the system and environment spin fully
polarized, the time dependent average of the system spin is given by
\[
p(t) = \<S \uparrow|R^{\dagger} \sigma_z(t) R|\uparrow S\> ~~,
\]
where the first index of the ket is the system state and the second is
the environment.  The spin operator $\sigma_z(t)$ is in the Heisenberg
representation, and its time evolution is governed by the
transformed Hamiltonian,
$H^{\prime}$.  We now introduce the Liouvillian time evolution
operator $L$, which is defined generically for a \ha $H$, Hilbert space
$\{|j\>\}$ and operator $O$ as follows:
\[
\<i| O(t)|j\> = \sum_{i^\prime,j^\prime}{(i j|e^{i L t}|i^\prime
j^\prime) \<i^\prime|O(0) |j^\prime\>} ~~.
\]
The time evolution operator $Q(t)\equiv \exp{(iLt)}$ is a fourth rank
operator whose state space vectors are denoted by a ket-parenthesis,
$|ij)$.  All dynamical information is contained in $Q$ and 
the Laplace transform of $Q$
\[
(j \mu k \nu|Q(z)|l \mu^\prime m \nu^\prime) \equiv (j \mu k
\nu|\frac{1}{z-iL}|l \mu^\prime m \nu^\prime) ~~,
\]
is the most convenient form to calculate.  For the initial conditions
specified above, it follows that the Laplace transform of $p(t)$ is
given by:
\begin{equation}
\label{time_ev}
p(z) = \sum_{m \mu j k}{\alpha_j \alpha_k^* (k \uparrow j \uparrow|
Q(z) |m \mu m \mu) \mu} \nonumber
\end{equation}
where the $\alpha$'s are the rotation group matrix elements, 
\[
\alpha_j \equiv \<j \uparrow| R |\uparrow S\>
\]

Following Dattagupta {\em et al.} \cite{dattagupta89}, we 
concentrate on the
``environment-averaged'' time evolution operator defined by performing
the environment sums in (\ref{time_ev})
\[
(\mu \nu |\<Q(z)\>|\mu^\prime \nu^\prime) \equiv \sum_{kjm}
{\alpha_j \alpha_k^* (k \mu j \nu| Q(z) |m \mu^\prime m \nu^\prime)}
\]
The Liouvillian is now split into perturbation ($L_i$) and environment
($L_e$) parts, corresponding to the two terms in equation
(\ref{ham3}). Then the environment averaged time development operator
may be computed perturbatively as
\begin{equation}
\label{av_q}
(\mu \nu |\<Q(z)\>|\mu^\prime \nu^\prime) = (\mu \nu |
\frac{1}{z-\<M\>_c} | \mu^\prime \nu^\prime)~~,
\end{equation}
where
\begin{eqnarray}
\label{selfenergy}
(\mu \nu |\<M\>_c | \mu^\prime \nu^\prime)=&& 
z(\mu \nu |\<i L_i G\> | \mu^\prime \nu^\prime)\\
-&&z(\mu \nu |\<L_i G L_i G\> | \mu^\prime \nu^\prime) \nonumber \\
+&&z(\mu \nu |\<L_i G\>^2 | \mu^\prime \nu^\prime)+ \ldots ~~,\nonumber
\end{eqnarray}
and $G$ is the Liouvillian propagator for the environment degrees of
freedom, $G^{-1}=z-iL_e$.

\begin{figure}[tb]
\epsfxsize=3.5in
\centerline{\epsfbox{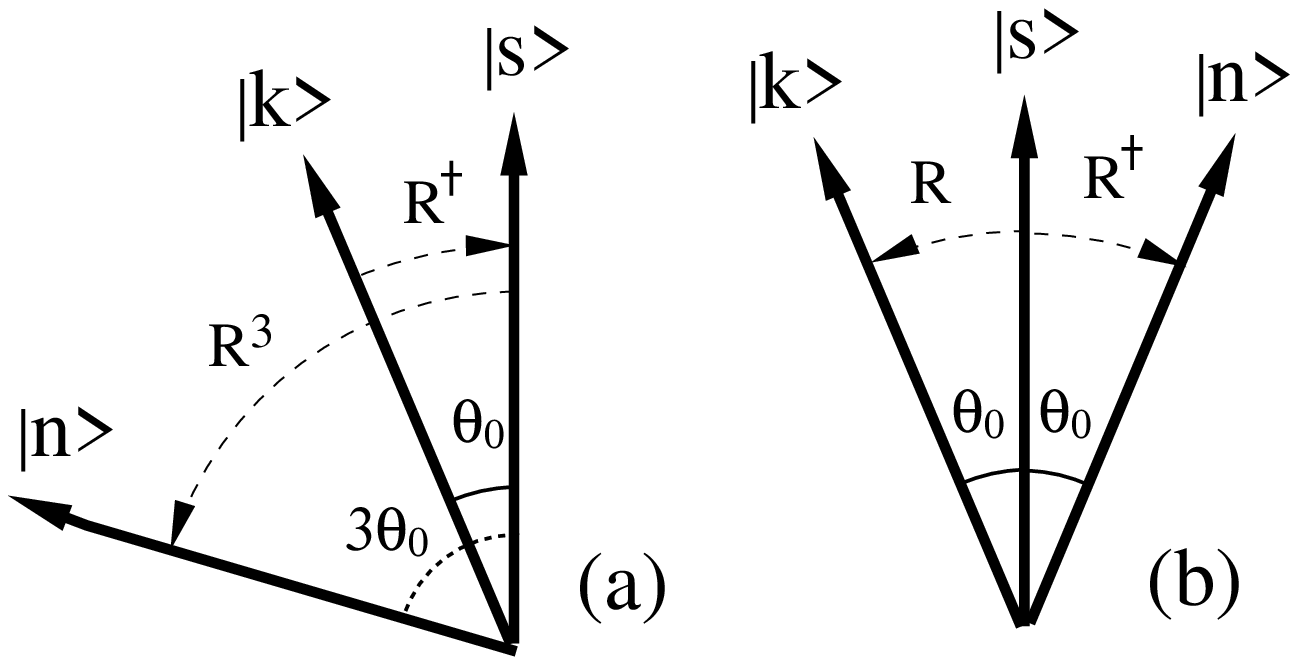}}
\vskip 0.5truein
\protect\caption{Fluctuation matrix elements depicted for processes
contributing to $B(z)$ and $C(z)$.}
\label{fluctuation}
\end{figure}

We can now calculate the above terms. 
The results through second order in $\Delta$ for the three
self-energies in the order appearing in equation (\ref{selfenergy})
(notated $M_1$, $M_{21}$, $M_{22}$) are given below:
\begin{equation}
\label{m1}
M_1 = i\Delta\<S|R^2|S\> (\delta \otimes \sigma^x - \sigma^x \otimes
\delta)
\end{equation}
\begin{eqnarray}
\label{m21}
M_{21} = &&-\frac{1}{2}B(z)\delta \otimes (\delta + \sigma^z) -
\frac{1}{2}B^*(z)(\delta + \sigma^z) \otimes \delta \nonumber \\
&&-\frac{1}{2}C(z)\delta \otimes (\delta - \sigma^z) - 
\frac{1}{2}C^*(z)(\delta - \sigma^z) \otimes \delta \nonumber \\
&&+\frac{1}{2}(B(z)+B^*(z))S^- \otimes S^-  \\
&&+\frac{1}{2}(C(z)+C^*(z)) S^+ \otimes S^+
\nonumber
\end{eqnarray}
\begin{equation}
\label{m22}
M_{22} = \half \frac{\Delta^2}{z} |\<S|R^2|S\>|^2 (\delta \otimes \delta
- \sigma^x \otimes \sigma^x)
\end{equation}
The outer product notation $\alpha \otimes \beta$ corresponds to
$\alpha_{\mu \mu^\prime} \beta_{\nu \nu^\prime}$. The $z$ dependences
of $M_{21}$ are given by
\begin{eqnarray}
B(z) =&& \frac{\Delta^2}{4}\sum_{kn}{\frac{1}{z-i\epsilon_{kn}}\<n|R^3|S\>
\<S|R^\dagger|k\>\<k|R^{\dagger 2}|n\>} \nonumber \\
C(z) =&& \frac{\Delta^2}{4}\sum_{kn}{\frac{1}{z-i\epsilon_{kn}}
\<n|R^\dagger|S\>\<S|R^\dagger|k\>\<k|R^2|n\>} \nonumber
\end{eqnarray}
where $\epsilon_{kn} \equiv \epsilon_k - \epsilon_n$ is the bare
energy difference.  The fluctuation matrix elements are depicted
graphically in Fig. (\ref{fluctuation}). Since $\epsilon_k -
\epsilon_n$ depends only upon $k-n$, $n \equiv k+r$, may be written in
terms of a relative index $r$ and the $k$ sum performed. For large
$S$, the matrix elements, as functions of $r$, are peaked at the
most probable values of $r$. Now, the two self energies may be written
in the large $S$, continuum limit as follows:
\begin{eqnarray}
{\rm Re} C(z) =&& \frac{\Delta^2}{2 \lambda \sqrt{\pi}}
\int_{-\infty}^{\infty}{d \epsilon \frac{z}{z^2 + \epsilon^2}
e^{-(\frac{\epsilon}{\lambda})^2}} \nonumber \\
{\rm Re} B(z) =&& \frac{\Delta^2}{2 \lambda \sqrt{\pi}}
\int_{-\infty}^{\infty}{d \epsilon \frac{z}{z^2 + \epsilon^2}
e^{-(\frac{\epsilon-\overline{\epsilon}}{\lambda})^2}} \nonumber
\end{eqnarray}

These self energies cannot be calculated in closed form; however, they
can be evaluated in several limits that are physically significant.
For $C(z)$ we find,
\begin{eqnarray*}
{\rm Re} C(z) \sim &&\frac{\Delta^2}{\lambda} \,\,\,\,\,\,  z<<\lambda \\
{\rm Re} C(z) \sim &&\frac{\Delta^2}{z} \,\,\,\,\,\,\,\, z >> \lambda \,\,\,{\rm
or}\,\,\, \lambda << \omega
\end{eqnarray*}
and for $B(z)$,
\begin{eqnarray*}
{\rm Re} B(z) \sim
&&\frac{\Delta^2}{\lambda}e^{\overline{\epsilon}^2/\lambda^2}
\,\,\,\,\,\,\,\, \,z<<\lambda \\ {\rm Re} B(z) \sim &&\Delta^2 \frac{z}{z^2
+ \overline{\epsilon}^2} \,\,\,\,\,\, \lambda << \omega \\ {\rm Re}
B(z) \sim &&\frac{\Delta^2}{z}
\,\,\,\,\,\,\,\,\,\,\,\,\,\,\,\,\,\,\,\,\,\, z >> \lambda
\end{eqnarray*}
where $\overline{\epsilon}=\omega(\cos{\theta_0}-\cos{3\theta_0})$.

To compute the Laplace transform of $p(t)$ the self energies (\ref{m1}),
(\ref{m21}), and (\ref{m22}) must be inserted in eqn. (\ref{av_q}) and the
fourth rank operator inverted. The general form is complicated,
but for small $z$, the analytic structure reduces to
\[
p(z) = \frac{{\rm Re} C(z) + z}{z(z+ {\rm Re} C(z) + {\rm Re} B(z))} ~~.
\]
In the limit of large environmental coupling, the real pole dominates
and the behavior for $p(t)$ is exponential decay; at zero
temperature, the system prepared in one state will leak into the other
state. Thus, we get,
\begin{equation}
\label{decay}
p(t) = e^{-t \Delta^2/\lambda} \,\,\,\,\,\,\, \lambda >> \Delta
\end{equation}
For small environmental coupling,
the pole in $p(z)$ is imaginary and $p(t)$
oscillates, given by 
\begin{equation}
\label{oscillation}
p(t) = \cos{\Delta t} \,\,\,\,\,\,\, \lambda << \Delta
\end{equation}
The results Eq. (\ref{decay}) and (\ref{oscillation}) summarize the
two extremal behaviors of $p(t)$, but give no information about the
critical point between.

\section{Numerical results for the Mermin model}
\label{num_res}

There are two main differences between the predictions of the
Liouvillian perturbative scheme and the exact numerical behavior.
First, the Liouvillian formalism does not capture the phase transition
observed in finite size scaling of the numerical results (as well as
in the mean field solution) described below. 
This result is interesting, given that the  
Liouvillian formalism does describe the phase transition at $\alpha=1$ of
the spin-boson \cite{dattagupta89} and spin-spin bath \cite{hanggi98}
models. Consequently, one would expect it to work for the simpler 
monochromatic
spectrum of the Mermin model.  Secondly, the Liouvillian formalism
predicts exponential decay in $p(t) \sim e^{-t \Delta^2/\lambda}$ for
large environmental coupling $\lambda$. The corresponding spectral
function is Lorentzian centered at $\nu =0$, similar to that of the
spin-boson model at the Toulouse point $\alpha= 1/2$.  However, in our
numerical results we will show that the spectral function has only a
finite energy $(\sim O(\Delta))$ feature and an exponentially small
frequency $(\sim O(\Delta e^{-N/2}))$ delta function feature (for
environmental coupling larger than the critical value.)

Let us now summarize briefly, our numerical findings. The transition to
decoherence is induced by strong coupling to environmental degrees of
freedom (large $\lambda$), or, equivalently, 
a small system energy scale (small $\Delta$).
Below a critical coupling, $\lambda_c$, all spectral weight of the
dynamical \sus \be \label{imchi} \imchi \equiv \mbox{Im}
\,\,\frac{i}{4}\int_{-\infty}^{\infty}{dt
\<0|[\sigma(t),\sigma(0)]|0\> \theta (t) e^{i\nu t}} \ee 
resides in the
principal resonance of the two level system
at an energy $\Delta$.  When the coupling is
larger than $\lambda_c$, a new exponentially small
energy scale $O(\Delta e^{-N/2})$ emerges, associated with a broken
symmetry, $\<\sigma\>\neq 0$, in the thermodynamic limit $N
\rightarrow \infty$.  This feature corresponds to tunneling modified
by a Franck-Condon type overlap factor.  As the coupling is increased,
spectral weight is shifted continuously to the ``near-zero'' frequency
channel.  The weight of the delta-function, $\delta(0^+)$, is simply
the order parameter, $|\<0|\sigma|0\>|^2$ (or, more exactly,
$|\<1|\sigma|0\>|^2$, for large but finite $N$). However, the
dynamical \sus obeys a sum rule, implying that an incompletely formed
broken symmetry state leaves some spectral weight at the position of
the principal resonance, set by $\Delta$.

To calculate the dynamical susceptibility we consider the response of
the system (\ref{ham2}) to an external field $h(t) =
h\cos{\nu t}$: \be
\label{ham_ext}
H_{\mbox{ext}} = -h \frac{\sigma_z}{2} \cos{\nu t}
\ee

The \ha (\ref{ham2}) was diagonalized and the complete order parameter
matrix, $\sigma_{ij} \equiv \<i|\sigma|j\>$, was computed.  It is
natural to work with the dimensionless coupling constant, $\kappa
\equiv 2\omega \Delta/\lambda^2$;  The critical coupling implied in
the $N \rightarrow \infty$ calculation is then $\kappa = \kappa_c =
1$.

\begin{figure}[tb]
\epsfxsize=3.5in
\centerline{\epsfbox{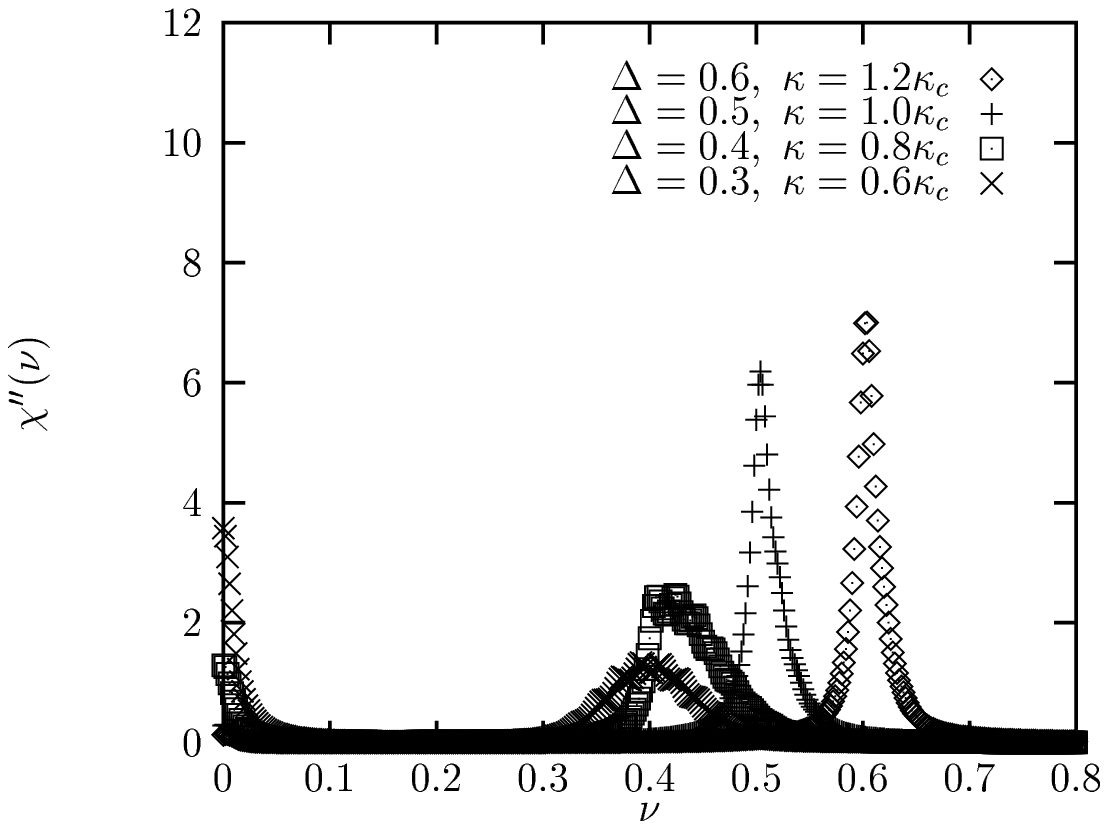}}
\protect\caption{Evolution of spectral weight in the Mermin Model.
$\imchi$ is shown for four values of $\kappa$ decreasing below the
critical value, $\kappa_c$.  Despite the change in the position of the
principal resonance (due to changing $\Delta$), spectral weight emerges
at $\nu \sim 0$ abruptly at $\kappa_c$ and the two features remain
distinct.}
\label{linear_sus}
\end{figure}

Fig. (\ref{linear_sus}) shows the behavior of $\imchi$ as the system
energy scale $\Delta$ is reduced, corresponding to a range $\kappa =
0.6\kappa_c - 1.2\kappa_c$.  These computations were performed for
$N=80$. Computations on larger spin environments (up to $N=300$)
suggest that there is little
change beyond $N=80$.  An artificial broadening ($\delta=0.01$) was
added to make the features more visible. As seen in this sequence,
$\kappa=\kappa_c$ is marked by the appearance of the Franck-Condon
resonance at an exponentially small energy scale ($\nu_{10} \sim
10^{-3}$).  The spectral weight remaining at $O(t)$ when $\kappa >
\kappa_c$ clearly exhibits inelastic broadening, although the
resonance essentially disappears before becoming critically damped.

The counterpart, at finite $N$, to $\<\sigma\> \neq 0$ at infinite $N$
is the spectral weight at the Franck-Condon resonance,
$\sigma_{10} \neq 0$.  Fig. 3 demonstrates that $\sigma_{10}$ behaves
as expected in a second order phase transition; the slope at $\kappa =
\kappa_c^+$ grows with increasing $N$.  The matrix, $\sigma_{ij}$, and
consequently, $\imchi$, obey a sum rule: 
\be 
\label{sum_rule}
\sum_i{|\sigma_{ij}|^2} = 1 \ee An incompletely developed broken
symmetry ($|\sigma_{10}|^2 < 1$) means that spectral weight remains at
the principal resonance at $O(\Delta)$.  
\begin{figure}[tb]
\epsfxsize=3.5in
\centerline{\epsfbox{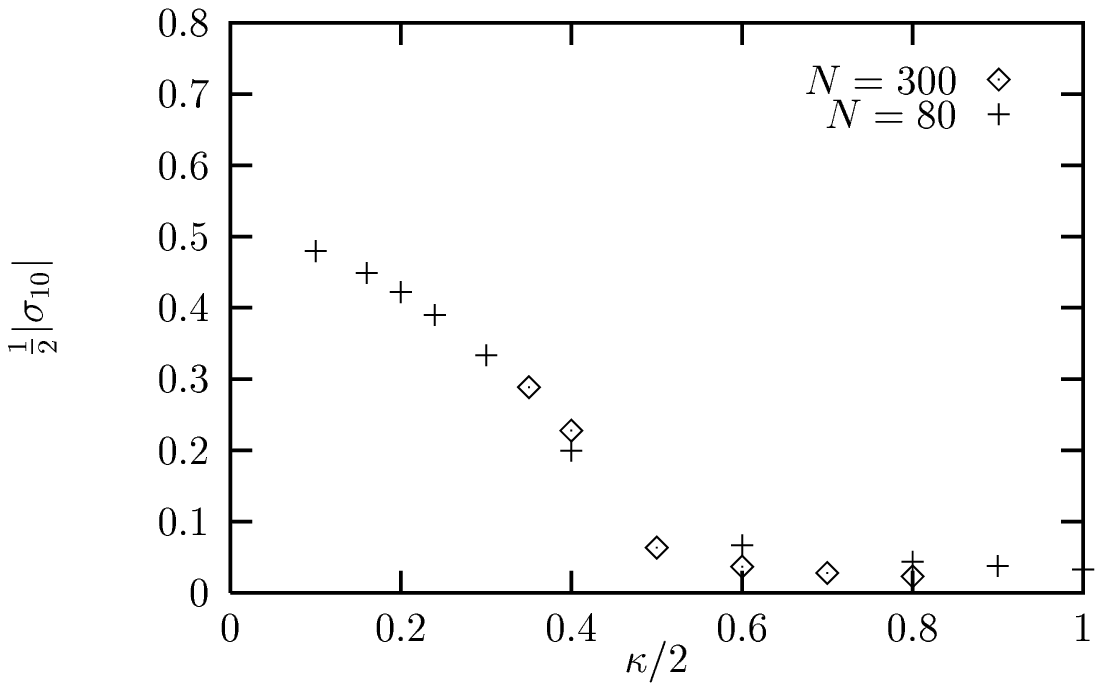}}
\protect\caption{The order parameter, $\sigma_{10}$,
is plotted as a function of $\kappa$ and exhibits second order phase
transition behavior.  $\kappa$ was varied by changing $\Delta$.  Comparison
of $N=80$ and $N=300$ data show the slope steepening at the critical
point.  When $\kappa > \kappa_c$, there is no weight in the
Franck-Condon resonance other than that attributed to the rounding of
the phase transition at finite $N$.}
\end{figure}

The quantum fluctuations missed by the mean field analysis but
captured in the numerics may be qualitatively divided into two types:
overlap effects and inelastic effects.  In the former type, the system
in the $|+\>$ state leaves an ``imprint'' upon the environment which,
being nearly orthogonal to the imprint left by system state
$|-\>$, reduces the tunneling amplitude by an exponentially small
factor. The new tunnel splitting is thus reduced by a Franck-Condon
type overlap factor: \be \Delta^* =
\Delta\<\theta_0|R_y(2\theta_0)|\theta_0\> = \Delta \cos^N{\theta_0} =
\Delta \frac{1}{(1+\lambda^2/\omega^2)^{N/2}} \ee where
$R_y(2\theta_0)$ is the rotation operator.  The resonance then shifts
from $O(\Delta)$ to a smaller energy scale $O(\Delta e^{-N/2})$. In the
inelastic type fluctuation, the system forces the environment to make
transitions to an excited state dissipating an amount energy
$\lambda^2/\omega$ per period. The resonance is then broadened by
$O(\lambda^2/\omega)$ but remains nominally at $O(\Delta)$.  

These results show that the spin dynamics of the Mermin model is 
significantly different from that of the SBH with an ohmic
bath. The former, as our results show, only exhibits overlap effects
at the critical coupling, but not before. In the SBH, the coupling to
the environment produces {\sl finite} overlap effects for all couplings
$0<\alpha<1/2$, and a Franck-Condon type reduction in energy scale,
given by:
\begin{equation}
\Delta_{\rm eff} = \Delta
(\frac{\Delta}{\omega_c})^{\frac{\alpha}{1-\alpha}}~~.
\end{equation}
At the same time, inelastic effects decohere the resonance leading to
the Toulouse limit (at $\alpha=1/2$), prior to localization 
(which occurs at
$\alpha=1$), and corresponding to a complete inelastic broadening of the
resonance. \cite{stockburger98,lesage96,costi_96}. 
In the Mermin model, the
quantum resonance of the system spin, although damped, remains
intact through the localization transition.

\section{Conclusion}
\label{conclusion}

In this paper we have compared the dynamical behavior of a quantum
dissipative system possessing a monochromatic spectrum (the Mermin
model) with that of the spin-boson model which possesses a continuous
bath spectrum.  We find an important difference between the dynamical
behavior of the two models, with the Mermin model reaching a localized
state, as environmental coupling is increased, without the primary
resonance becoming overdamped.  When the critical coupling is reached,
spectral weight is tranferred across a gap of $O(\Delta)$ to a zero
frequency delta function, indicating the presence of a localized
state.  In effect, the Mermin model exhibits no {\sl finite}
environmental overlap effects until an infinite one---the
orthogonality catastrophe---localizes the system.  In contrast, the
primary resonant feature in spin-boson model must soften to zero
frequency and become completely overdamped (the Toulouse point
$\alpha=1/2$) before a localized spectral feature appears at
$\alpha=1$.  Furthermore, the Liouvillian technique, which has been
successful in capturing the localization phase transition in the SBH
and related models fails for the Mermin model.  

The Mermin model is a prototype system for a weakly coupled $O(1/N)$
monochromatic spin bath, and judging from the results of Hanggi, a
weakly coupled monchromatic {\sl oscillator} bath is likely to exhibit
the same behavior. In contrast, the spin-boson model employs a
featureless, power-law spectrum of envrionmental oscillators and
exhibits universal behavior that depends only upon the exponent of the
spectrum.  This behavior, in the ohmic case, includes a localization
transition, but one with a qualitatively different character than that
of the Mermin model.  It is natural, then, to ask under what
conditions does a quantum environment become sufficiently
``featureless'' to fit into the universal framework of the spin-boson
model?

\acknowledgments

G.L. gratefully acknowledges the support of the Cottrell
Foundation through Research Corporation Grant number CC3834.  Work
performed at BNL supported by the U.S. DOE under contract no. DE-AC02
98CH10886. V.N.M. is supported by NSF Grant DMR-9104873.

\end{document}